\documentclass[12pt]{article}
\pdfoutput=1
\usepackage{amsmath,amssymb,amsfonts,setspace,graphicx,xcolor,cite,upgreek}
\usepackage[utf8]{inputenc}
\usepackage[hidelinks]{hyperref}
\usepackage{notes2bib}

\def\O{{\mathcal O}}
\newcommand{\be}{\begin{equation}}
\newcommand{\ee}{\end{equation}}
\newcommand{\bea}{\begin{eqnarray}}
\newcommand{\eea}{\end{eqnarray}}
\newcommand{\comment}[1]{}
\newcommand{\expect}[1]{\left\langle #1 \right\rangle}
\newcommand{\Cint}{C\kern-1em\int}
\def\d{\partial}
\def\ep{\epsilon}

\def\di{\mathrm{d}}

\def\Pcal{\mathcal{P}}
\def\Tr{\text{Tr}}

\def\multiset#1#2{\ensuremath{\left(\kern-.3em\left(\genfrac{}{}{0pt}{}{#1}{#2}\right)\kern-.3em\right)}}
\usepackage{accents}

\usepackage{mathtools}

\begin{document}
\begin{center}
{\bf \Large Low-temperature entropy in JT gravity}
\vskip 1cm
\textbf{Oliver Janssen$^{\star,\dagger}$} and \textbf{Mehrdad Mirbabayi$^\star$} \\
\vskip 0.4cm
{\em $^\star$International Centre for Theoretical Physics \\
$^\dagger$Institute for Fundamental Physics of the Universe \\
Trieste, Italy}
\vskip 0.7cm
\end{center}

\vspace{.8cm}

\noindent \textbf{Abstract:} 
For ensembles of Hamiltonians that fall under the Dyson classification of random matrices with $\upbeta \in \{1,2,4\}$, the low-temperature mean entropy can be shown to vanish as $\expect{S(T)}\sim \kappa T^{\upbeta+1}$. A similar relation holds for Altland-Zirnbauer ensembles. JT gravity has been shown to be dual to the double-scaling limit of a $\upbeta =2$ ensemble, with a classical eigenvalue density $\propto e^{S_0}\sqrt{E}$ when $0 < E \ll 1$. We use universal results about the distribution of the smallest eigenvalues in such ensembles to calculate $\kappa$ up to corrections that we argue are doubly exponentially small in $S_0$. 

\vspace{-0.7cm}

\setcounter{page}{1}
\vskip 1 cm
\section{Introduction}
Thermodynamic properties of a single system at temperature $T$ can be determined from the knowledge of its thermal partition function $Z(T) = \Tr \, e^{-H/T}$. For instance, the entropy is given by the derivative of the free energy $F(T)\equiv - T \log Z(T)$,
\be
	S(T) = - \d_T F (T) \,.
\ee
However, if we are dealing with an ensemble of disordered systems, then knowledge of the disordered average $\expect{Z(T)}$ is insufficient to determine $\expect{S(T)}$. Instead we need to know the {\em quenched free energy} \cite{Anderson}
\be
F_q(T) \equiv -T \expect{\log Z(T)} \,.
\ee
Typically, large groups of energy levels contribute collectively at high temperature and the distinction between various members of the ensemble smears out. In this limit the {\em annealed free energy}, 
\be
F_a(T) \equiv -T  \log \expect{Z(T)}\,,
\ee
is a good approximation to $F_q(T)$. The distinction between the annealed and quenched free energies is noticeable at low temperature.

This subtlety appears to have some relevance to gravitational physics \cite{Engelhardt}. Euclidean wormhole contributions to the gravitational path integral have led to novel holographic dualities in which gravity computes ensemble averages of boundary observables. In particular, a 2D model of gravity on negatively curved spacetime, known as Jackiw-Teitelboim or JT gravity \cite{Jackiw,Teitelboim}, has been shown to be dual to the double-scaling limit of a unitary ensemble \cite{Saad}. That is, the JT path integral with $n$ boundaries of lengths $T_1^{-1},T_2^{-1},\cdots,T_n^{-1}$ is shown to give
\be
	\Pcal_{\rm JT}(T_1^{-1},T_2^{-1},\cdots) = \expect{Z(T_1)Z(T_2)\cdots}_{\rm DS} \,,
\ee
where the expectation on the right is over an ensemble of random $N\times N$ Hermitian matrices
\be \label{matrix}
\expect{Z(T_1)Z(T_2)\cdots} = 
\int \di H ~ e^{-N \, \Tr \,  V(H)} ~ \text{Tr} \, e^{-H/T_1} ~ \text{Tr} \, e^{-H/T_2} \cdots \,,
\ee
and the subscript DS stands for the double-scaling limit. It corresponds to sending $N\to \infty$ while rescaling the parameters of $V(H)$ in a particular fashion, and focusing one's attention to the edge of the spectrum (as will be discussed further in section \ref{sec:review}). Several variants of this duality have been proposed in \cite{Stanford}.

At low temperature, when the boundary lengths grow, it becomes more favorable to connect the boundaries via wormholes. This transition has been identified in \cite{Okuyama_rsb,Okuyama,Okuyama_multi,Johnson_Tcr}. As expected for disordered systems, in this limit the non-factorization of $\expect{Z(T)^n}$ becomes significant and $F_a(T)$ and $F_q(T)$ deviate from one another. The recent work \cite{Engelhardt} made the interesting observation that the annealed free energy in JT is so off at low temperature that it predicts negative entropy. It would be interesting to find a prescription that directly computes $F_q$ on the gravity side -- one that applies to any gravitational theory that can be interpreted as a disordered average. See \cite{Engelhardt} for further comments.

Here we instead focus on the particular example of JT gravity and its cousins with known ensemble duals to analyze the low-temperature behavior of $F_q(T)$ on the matrix model side. Earlier works in this direction include \cite{Johnson,Okuyama:2020mhl,Okuyama21}. What underlies our analysis is the observation that, at sufficiently low temperature, $F_q$ is dominated by the distribution of the smallest eigenvalues of $H$ \cite{Okuyama:2020mhl}. In section \ref{sec:fq}, we see how this fixes the low-temperature scaling of $\expect{S(T)}$ for the ensemble dual to JT, as well as all other ensembles considered in \cite{Stanford}.

Matrix ensembles are known to have universal behaviors in the bulk and near the edge of the collective eigenvalue density (for an overview see \cite{kuijlaars}). In the case of JT, the edge region is controlled by the well-known Airy kernel of unitary ensembles (corresponding to the Dyson index $\upbeta = 2$), and much is known about the distribution of the smallest eigenvalue \cite{TW}, as well as its distance to the next eigenvalue \cite{WBF13,PS}. Using these results, in section \ref{sec:airy}, we calculate 
\be
	F_q^{\rm Airy}(T)=(1.77\cdots)\times 2^{-1/3} e^{-2S_0/3} - \frac{7\pi^4}{360 }e^{2 S_0}T^4 +\O(T^{6}) \,,
\ee
where $e^{-2S_0}$ is the genus counting parameter of JT. In section \ref{sec:dis} we estimate the difference between the JT and Airy results, concluding that the relative corrections are doubly \\ exponentially suppressed in $S_0$.

\section{Review of JT gravity and its matrix dual}\label{sec:review}
JT gravity is a simple dilaton-gravity model in two spacetime dimensions \cite{Jackiw,Teitelboim}. Its bulk action with an appropriately rescaled negative cosmological constant is
\be
	I_{\rm JT} = -\frac{1}{2} \int \di ^2 x \sqrt{g} \phi(R+2) - \frac{S_0}{4\pi} \int \di^2 x\sqrt{g} R \,,
\ee
where $R$ is the scalar curvature of the metric. The Einstein-Hilbert term (together with the appropriate Gibbons-Hawking boundary term) is topological in 2D, equal to the Euler characteristic of the manifold, and leads to a suppression of geometries with higher genus and with more boundaries. Hence, the partition function with $n$ boundaries can be formally expanded as $Z_n = \sum_{g\geq 0} e^{(2-2g-n) S_0} Z_{g,n}$. In particular, the disk partition function is given by \cite{MSY,Stanford_Z0}
\be
	Z_{0,1}(T) = \frac{T^{3/2} e^{2\pi^2 T}}{\sqrt{2\pi}} \,,
\ee
which defines a genus-zero ``density of states'' via $Z_{0,1}(T) = \int_0^\infty \di E \rho_0(E) e^{-E/T}$,
\be \label{rho0}
\rho_0(E) \equiv \frac{1}{2\pi^2}\sinh(2\pi\sqrt{2 E}) \,, \qquad E>0 \,.
\ee
The matrix model dual to JT gravity is uniquely fixed (at the perturbative level) by this data \cite{Saad}. Below, we will summarize some of its key properties (a review on matrix models can be found in \cite{Eynard}). 

The first hint for the duality comes from the fact that matrix integrals of the form \eqref{matrix} also admit a genus expansion, in $1/N$, upon using the double-line formalism \cite{tHooft,Brezin}. For analytic potentials $V(H)$ and observables such as $Z(T)$, which depend only on traces of powers of $H$, the integrand is solely a function of the eigenvalues of $H$ after diagonalization. The Jacobian of the transformation $H = U^\dagger {\rm diag}(\lambda_1,\lambda_2,\cdots ) U$ is a Vandermonde determinant, giving the following partition function for the matrix eigenvalues
\be \label{Zm}
\mathcal{Z} = \int \{\di \lambda_i\}  \prod_{i<j} (\lambda_i-\lambda_j)^2 \prod_i e^{-N V(\lambda_i)} \,, \qquad -\infty<\lambda_i<\infty \,.
\ee
As reviewed in \cite{Stanford}, if the random Hamiltonians are invariant under a time-reversal ${\sf T}$, then depending on whether ${\sf T}^2 = 1$ or $-1$, we will have instead a Jacobian $\prod_{i<j}|\lambda_i-\lambda_j|^\upbeta$ with $\upbeta = 1$ or $4$. The variants of JT gravity considered in \cite{Stanford} cover all these choices of $\upbeta$ (known as Dyson ensembles \cite{Dyson}), as well as the seven Altland-Zirnbauer (AZ) ensembles \cite{AZ}, that are defined on semi-infinite intervals, and have an additional parameter $\upalpha\in\{0,1,2,3\}$ 
\be \label{ZAZ}
\mathcal{Z}_{\rm AZ} = \int \{\di \lambda_i\}  \prod_{i<j} |\lambda_i-\lambda_j|^\upbeta \prod_i \lambda_i^{\frac{\upalpha-1}{2}} e^{-N V(\lambda_i)},\qquad {\lambda_i>0} \,.
\ee
As we will see, the structure of the low-temperature expansion of the quenched free energy is fixed in terms of $\upbeta$ (and $\upalpha$), but our computation of the coefficients is restricted to the Dyson ensemble with $\upbeta = 2$.

The Vandermonde determinant acts as a repulsive force among the eigenvalues. This manifests itself in the expectation value of the eigenvalue density 
\be
	\rho^{\rm total}(\lambda) = \sum_i \delta(\lambda-\lambda_i) \,,
\ee
where the superscript total means that this density times $\di \lambda$ gives the actual number of eigenvalues in that interval. For instance, in a Gaussian unitary ensemble with $V(H) = \frac{2}{a^2} H^2$, the leading perturbative answer for the expected value of $\rho^{\rm total}$ at large $N$, $\rho_0^{\rm total}$, is the famous Wigner semicircle 
\be \label{wigner}
\expect{\rho^{\rm total}(\lambda)} = \frac{2 N}{\pi a^2} \sqrt{a^2 - \lambda^2}+\O(N^{-1}) \,, \qquad  |\lambda|<a \,,
\ee
which extends well beyond the width of the Gaussian factor $a/\sqrt{N}$ in \eqref{Zm} because of the eigenvalue repulsion. Other choices of $V$ would lead to different genus-zero or ``global" densities, so one could take $\rho_0^{\rm total}$ rather than $V$ as the definition of the model. However, unless the potential is fine-tuned, the behavior near the edge of the distribution is universal in the $N\to \infty$ limit. To focus on this region, one takes the double-scaling limit. In the Gaussian example, we take
\be
	\lambda = a + x \,, \qquad a,N \to \infty \qquad\text{with}\qquad  N\left(\frac{2}{a}\right)^{3/2} = 1 \,,
\ee
to obtain the genus-zero density in the allowed region $x<0$
\be\label{sqrt}
\rho_0^{\rm total} (x)  = \frac{1}{\pi}\sqrt{-x} \,.
\ee
(There is a reflection symmetry between the upper edge and the lower edge of \eqref{wigner}. It would be more physical to focus on the lower edge, i.e. on the low-energy spectrum, but to comply with the math literature we focus on the upper edge and apply the reflection when comparing with JT.)

The universality of \eqref{sqrt} is a consequence of the fact that unless the relative scaling of various terms in $V$ are fine-tuned as $N \to \infty$, the nontrivial features of $\rho_0$ are sent to infinity. Several exact results are known in this limit, using the method of orthogonal polynomials and the resulting Airy kernel.

In the case of JT gravity, we see that its genus-zero total density $e^{S_0} \rho_0(E)$ approaches \eqref{sqrt} if we identify
\be\label{x(E)}
x =- 2^{1/3} e^{2 S_0/3} E \,,
\ee
and take $E\ll 1$. Put differently, there is an ``Airy limit'' of JT gravity corresponding to $e^{S_0}\to \infty$ with $x$ kept finite. The deviation between \eqref{rho0} and \eqref{sqrt} at finite $S_0$ results from a carefully designed potential $V$ when taking the double-scaling limit. The explicit form of this potential is not needed. The knowledge of $\rho_0$ (or the closely related spectral curve) is enough to set up the matrix model genus expansion, and this was shown in \cite{Saad} to match the genus expansion of JT gravity.

\section{Low-temperature free energy in matrix models}\label{sec:fq}
In terms of the joint probability distribution $p(\lambda_0,\lambda_1,\cdots)$ of the matrix eigenvalues, which we order $\lambda_0<\lambda_1<\cdots$, the quenched free energy is given by
\bea
F_q(T) &=& -T \int \{\di\lambda_i\} \, p(\{\lambda_i\}) \log\left(\sum_{j\geq 0} e^{-\lambda_j/T}\right) \notag \\ [10pt]
&=&-T \int \{\di\lambda_i\} \, p(\{\lambda_i\}) \left[-\frac{\lambda_0}{T} +\log\left(1+\sum_{j\geq 1} e^{-\Delta_{j0}/T}\right)\right] \,, \label{fexpand}
\eea
where $\Delta_{ij} \equiv \lambda_i - \lambda_j$. Suppose we lower the temperature well below the typical spacing of eigenvalues near the edge of distribution. In JT gravity, this typical distance can be inferred from \eqref{rho0},
\be
	\rho^{\rm total}_0(\Delta_{\rm typ}) \Delta_{\rm typ} \approx 1 \Rightarrow \Delta_{\rm typ} \approx e^{-2 S_0/3} \,.
\ee
Then the $j^{\text{th}}$ term inside the log in \eqref{fexpand} is suppressed unless $\lambda_1,\lambda_2,\cdots,\lambda_j$ are all squeezed closer than $\Delta_{\rm typ}$ to $\lambda_0$. As a result, in Dyson ensembles with parameter $\upbeta$
\be\label{expmean}
\expect{e^{-\Delta_{j0}/T}} \underset{T\to 0}\propto T^{j\left(1+\frac{\upbeta}{2}(j+1)\right)} \,,
\ee
which can be verified as follows. First, changing the integration variables $\{\lambda_1,\cdots,\lambda_j\}\to\{\Delta_{10},\cdots,\Delta_{j0}\}$ gives
\be
\expect{e^{-\Delta_{j0}/T}} = \int_{-\infty}^\infty \di\lambda_0\left(\prod_{i=1}^j \int_{\Delta_{i-1 \, 0}}^\infty \di\Delta_{i0}\right)\left(\prod_{k=j+1}^{N-1}\int_{\lambda_{k-1}}^\infty \di\lambda_k\right) e^{-\Delta_{j0}/T} p(\{\lambda_i\}) \,,
\ee
where $\Delta_{0 0} \equiv 0$. Then rescaling $\Delta_{i0}\to T x_i$ gives $j$ factors of $T$ from the measure, while (defining $x_0\equiv 0$) the Vandermonde determinant contains a factor 
\be
|\Delta_{mn}|^\upbeta = T^\upbeta |x_m -x_n|^\upbeta
\ee
for each pair in $\{0, 1,\cdots,j\}$. This gives $\upbeta j(j+1)/2$ factors of $T$. Setting $T=0$ elsewhere results in a convergent integral. So the leading term as $T \rightarrow 0$ is as written in \eqref{expmean}.

In AZ ensembles with $\upalpha=0$, there is an extra contribution since after the same change of variables the $\lambda_0$ integral becomes
\be
\int_{0}^\infty d\lambda_0 \, \lambda_0^{-\frac{1}{2}}\left(\prod_{i=1}^j (\lambda_0 + T x_i)^{-\frac{1}{2}}\right) \times (\text{finite at $\lambda_0=0$}).
\ee
In the $T\to 0$ limit, this diverges as $T^{1-(j+1)/2}$ when $j>1$ and logarithmically when $j=1$. So the result for AZ ensembles with any $\upalpha$ can be written as
\be\label{expmeanAZ}
\expect{e^{-\Delta_{j0}/T}} \underset{T\to 0}\propto T^{j\left(1+\frac{\upbeta}{2}(j+1)\right)} \times \left\{\begin{array}{cc} 
T^{{\rm min}\left(0,1+(j+1)(\upalpha-1)/2\right)} &\quad j > 1 \\ [10pt]
1+ \delta_{\upalpha 0} \log\frac{\Delta_{\rm typ}}{T}  &\quad j = 1 \,. \end{array}\right.
\ee
Using these estimates, we can write \eqref{fexpand} at low temperature as
\be\label{Flead}
F_q(T) = \expect{\lambda_0}-T \expect{\log\left(1+e^{-\Delta_{10}/T}\right)}
+\left\{\begin{array}{cc} \O(T^{3(1+\upbeta)}) &\quad  \text{Dyson \& AZ with $\upalpha\neq 0$}\\[10pt]
\O(T^{3(1+\upbeta)-\frac{1}{2}})&\quad \text{AZ with $\upalpha = 0$.}\end{array}\right.
\ee
The low-temperature scaling of the second term on the RHS is the same as \eqref{expmean} or \eqref{expmeanAZ} with $j=1$, from which follows
\be
\label{S0}
	\expect{S(T)}\underset{T\to 0} \sim \kappa \, T^{1+\upbeta} \qquad \left[\times \log\frac{\Delta_{\rm typ}}{T} ~~ \text{in AZ with $\upalpha =0$}\right]
\ee
with $\kappa > 0$ an ensemble-dependent constant. This equation and Eq. \eqref{Flead} generalize the result of \cite{Okuyama:2020mhl} to all Dyson and AZ ensembles.

We see that the quenched free energy is guaranteed to give a positive averaged entropy. To explicitly calculate $F_q(0)$ and $\kappa$, we need to know the distribution of the smallest eigenvalue $p(\lambda_0)$, and the distribution of the first gap $p_{\rm gap}(\Delta_{10})$, or more precisely, the leading coefficient in its expansion $p_{\rm gap}(\Delta_{10}) \underset{\Delta_{10}\to 0}\sim \gamma (\Delta_{10})^\upbeta$ (accordingly, $\gamma (\Delta_{10})^{\upbeta}\log(\Delta_{\rm typ}/\Delta_{10})$ in AZ ensembles with $\upalpha = 0$). This last number $\gamma$ is not known in a generic ensemble. However, it is known exactly in the Airy limit, which applies at the edge of the spectrum in the $\upbeta = 2$ Dyson ensembles (i.e. of the type \eqref{Zm}). Below we will first use these results to compute $F_q^{\rm Airy}(0)$ and $\kappa^{\rm Airy}$, and then discuss how well they approximate JT gravity.

\section{The Airy limit}\label{sec:airy}
Consider a unitary ($\upbeta = 2$) Dyson ensemble with genus-zero density given by \eqref{sqrt}. The average density of eigenvalues in such an ensemble is given exactly by
\be\label{rhoA}
\expect{\rho^{\rm total}(x)} = {\rm Ai}'(x)^2 - x {\rm Ai}(x)^2 \,.
\ee
In particular, there is a nonzero but small probability of finding eigenvalues beyond the classical edge, i.e. $x>0$. The distribution of the largest eigenvalue in this ensemble has been found by Tracy and Widom \cite{TW} (a simple derivation can be found in \cite{Nadal}). The probability that $\lambda_{\rm max}<s$ is
\be
	F_2(s) = \exp\left(-\int_s^{\infty}(u-s)q^2(u) \di u \right) \,,
\ee
where $q$ is the solution to the following Painlev\'e II equation
\be \label{qofs}
q''(s) = 2 q^3(s)+ s \, q(s) \,, \qquad \text{with}\quad q(s) \underset{s\to \infty}\sim {\rm Ai}(s) \,.
\ee
The PDF of $\lambda_{\rm max}$ is given by 
\be\label{pdfTW}
p_{\rm TW}(s) =F_2'(s)= R(s)F_2(s) \,, \qquad R(s)\equiv \int_s^\infty q^2(u)\di u \,.
\ee
Asymptotically \cite{TW,Deift}
\be\label{asymptote}
p_{\rm TW}(s) \sim \left\{ \begin{array}{cc} \frac{1}{8\pi s} e^{-\frac{4}{3} s^{3/2}} \,, & \qquad s \to \infty\\
\frac{\tau_2}{4}(-s)^{15/8} e^{s^3/12} \,, & \qquad s \to -\infty \end{array} \right. \,.
\ee
where $\tau_2 = 2^{1/24}e^{\zeta'(-1)}$ ($\zeta$ is the Riemann zeta function). The right asymptotic, i.e. far in the forbidden region, coincides with the asymptotic behavior of the Airy density \eqref{rhoA}. This is because an eigenvalue found in the forbidden region is exponentially more likely to be the largest eigenvalue than any other one. The left asymptotic $s\to -\infty$ is steeper, which can be understood from the fact that $\O(|s|^{3/2})$ eigenvalues have to be significantly displaced.

The mean of $p_{\rm TW}$ can be evaluated numerically, $\langle \lambda_{\rm max} \rangle_{\upbeta = 2} = -1.77\cdots$. It determines the intercept $F_q(0)$, i.e the leading term in \eqref{Flead}, in the Airy limit. In the orthogonal and symplectic ensembles, $\upbeta = 1,4$ respectively, the distribution of the largest (or smallest) eigenvalue can be written in terms of $p_\text{TW}$ \cite{tracy1996}, and there we have $\langle \lambda_{\rm max} \rangle_{\upbeta = 1} = -1.21\cdots$, $\langle \lambda_{\rm max} \rangle_{\upbeta = 4} = -2.31\cdots$. We comment on the Altland-Zirnbauer ensembles in \S\ref{sec:dis}.

Calculating the subleading term in \eqref{Flead} requires knowledge of the distribution of the gap between the smallest two eigenvalues (the largest two in terms of the variable $x$). This distribution and importantly its asymptotic limits have been calculated by Perret and Schehr \cite{PS} in the Airy case. Denoting this gap by $\tilde r > 0$, combining equations (92) and (93) of \cite{PS} yields
\begin{align}
	p_{\rm PS}(\tilde r) &= \int_{-\infty}^\infty p_{\rm gap}(\tilde r|s) \, p_{\rm TW}(s) \di s \,, \label{pPS} \\
	p_{\rm gap}(\tilde r|s) &= \int_s^\infty f^2(-\tilde r,u)\di u - \frac{1}{R(s)} \left(\int_s^\infty f(-\tilde r,u)q(u) \di u\right)^2 \,, \label{cond2}
\end{align}
where $f$ is the solution to
\be
	\d_y^2 f(x,y) -[y+2 q^2(y)] f(x,y) = -x f(x,y),\qquad f(x,y)\underset{y\to \infty}\sim  {\rm Ai}(y-x) \,.
\ee
$p_{\rm gap}(\tilde r|s)$ can be thought of as describing the distribution of the largest eigenvalue in a new ensemble whose potential is related to the original potential by (see appendix \hyperref[app:wkb]{A2})
\be\label{V11}
	\tilde V(x) = V(x) - \frac{1}{N}\log(s-x)^2 \,, \qquad x<s \,.
\ee
In the low-temperature limit, we are sensitive to the small $\tilde r$ behavior of the distribution \eqref{pPS}. This is derived in \cite{PS} to be
\be\label{PS2}
p_{\rm PS}(\tilde r)= c_2 \tilde{r}^2 + \O(\tilde r^4) \,,
\ee
where $c_2 = 1/2$ and the coefficients of the higher order terms (only even powers appear) may be found algorithmically. Using this result and taking into account the rescaling \eqref{x(E)}, we find in the Airy limit
\begin{align}
	F^{\rm Airy}_q(T) &= - 2^{-1/3} e^{-2 S_0/3} \langle \lambda_\text{max} \rangle_\text{TW} - T \int_0^\infty \di \tilde{r} ~ p_\text{PS}(\tilde{r}) \, \log \left( 1 + e^{-\tilde{r}/\tilde{T}} \right) + \mathcal{O}(T^9) \notag \\[10pt]
	&= \left( 1.77 \cdots \right) \times 2^{-1/3} e^{-2S_0/3} - \frac{7 \pi^4}{360} e^{2 S_0} T^4 + \mathcal{O}(T^6) \label{FA} \,, 
\end{align}
where in the first line we defined $\tilde T = 2^{1/3} e^{2S_0/3}T$ (the combination that remains fixed in the Airy limit), and in the second line kept just the leading term in the small $\tilde r$ expansion \eqref{PS2}. In $\upbeta = 1,4$ ensembles the first subleading term would be of order $T^3$ and $T^6$ respectively, but we could not calculate the coefficients in these cases because the analog of Eq. \eqref{PS2} is not known there to our knowledge.

Before moving on to the comparison with JT, it is worth considering the asymptotics of $p_{\rm gap}(\tilde r|s)$. Expanding at small $\tilde r$ gives
\be\label{d2}
p_{\rm gap}(\tilde r|s) = d_2(s) \tilde r^2 + \mathcal{O}(\tilde r^3) ~~~~ \text{as } \tilde{r} \to 0 \,.
\ee
We will find in the appendix that 
\be
d_2(s) \sim \left\{ \begin{array}{cc} \frac{1}{128 \pi s^3} e^{-\frac{4}{3}s^{3/2}} \,, & \qquad s \to \infty \,, \\
\frac{1}{768} s^6 \,, & \qquad s \to -\infty \,. \end{array} \right.
\ee
We will also show how the WKB approximation in the potential \eqref{V11} reproduces the $s\to \infty$ behavior, and verify that unlike $p_{\rm PS}(\tilde r)$ the conditional probability $p_{\rm gap}(\tilde r|s)$ is not an even function of $\tilde r$.

\section{Discussion}\label{sec:dis}
We have seen that the low-temperature behavior of the free energy in matrix ensembles is fixed in terms of $\upbeta$ (and $\upalpha$) parameters. We expect the Airy result \eqref{FA} to provide an approximation for JT gravity at finite but large $S_0$ since both are matrix models with $\upbeta =2$ and a similar density near the edge of the distribution. Below we will give a heuristic estimate of the size of the corrections. 

First consider the distribution of the smallest eigenvalue $p_{\rm JT}(\lambda_0)$. After the change of variable \eqref{x(E)}, we expect this to agree well with $p_{\rm TW}(s)$ near $s=0$ but to deviate significantly for $|s|\gg s_*$, with $s_*\to\infty$ as $S_0\to \infty$. The thicker tail of $p_{\rm TW}(s)$, i.e $s\to \infty$ in \eqref{asymptote}, is expected to be more relevant for the estimate of the error. As discussed below \eqref{asymptote}, this tail of the distribution coincides with $\expect{\rho^{\rm total}(x)}$, which can in turn be evaluated using the WKB approximation. The WKB exponent is 
\be\label{exponent}
\log \expect{\rho_{\rm WKB}(s)} \approx 2N \int_0^s y(x) \di x \,,
\ee
where $y(x)$ is the spectral curve obtained by the analytic continuation of $i\pi\rho_0(x)$ to positive $x$ (for a derivation see e.g. \cite{Saad}). In the double-scaling limit, $\rho_0^{\rm total}= N \rho_0$ is kept finite. Noting that for JT gravity after the change of variable \eqref{x(E)}
\be
\rho_0^{\rm total}(x) = \frac{1}{\pi}\sqrt{-x} + \frac{2^{5/3}\pi}{3}e^{-2 S_0/3} |x|^{3/2} + \cdots \,,
\ee
the first correction to the WKB exponent \eqref{exponent} becomes $\O(1)$ when\footnote{We thank Douglas Stanford for the argument leading to this estimate.}
\be\label{s*}
s_* =\O(e^{4 S_0/15}) \,.
\ee
At this point $\expect{\rho_{\rm WKB}(s)}$ and hence the Tracy-Widom distribution is suppressed by 
\be\label{error}
p_{\rm TW}(s_*) = \O(e^{-\# e^{2S_0/5}}) \,.
\ee
We don't know how $p_{\rm JT}$ behaves far beyond this point. Assuming that it continues to decay (under a reasonable nonperturbative completion of the model) the error in approximating $F_q^{\rm JT}(0)$ with $F_q^{\rm Airy}(0)$ would be of the same order as \eqref{error}.

As for the coefficient of the $T^4$ term in \eqref{FA}, difference between JT and the Airy limit prediction can only result from the fact that the coefficient $c_2$ in \eqref{PS2} is not exactly $1/2$ in JT. As we saw, $c_2$ can be obtained by integrating $d_2(s)$ against the distribution of the largest eigenvalue, where $d_2(s)$ is the coefficient of the small gap expansion \eqref{d2} of the conditional probability distribution $p_{\rm gap}(\tilde r|s)$. Since $d_2(s)$ in the Airy limit grows at most like a power of $s$ at large $|s|$ while $p_{\rm TW}(s)$ decays exponentially, we expect $c_2^{\rm JT} -\frac{1}{2}$ to be of the same order as \eqref{error}. Hence our prediction for the JT entropy is
\be\label{S}
\expect{S(T)}_{\rm JT} = \frac{7\pi^4}{90} e^{2 S_0} T^3 \left( 1+\O(e^{-\#e^{2 S_0/5}}) \right) +\O(T^{5}) ~~~~~~ \text{as } T \rightarrow 0 \,.
\ee
It is interesting to contrast the Airy limit approximation for the JT free energy and entropy, which as we argued remain valid at arbitrarily small $T$, with its prediction for $\expect{Z(T)}$. In the latter case the Airy approximation breaks down for exponentially small $T<T_*$. This can be seen by first noting that in the small $T$ limit
\be
\expect{Z(T)} = \int_{-\infty}^\infty \di E \expect{\rho(E)} e^{-E/T}
\ee
is dominated by the small probability of finding an eigenvalue in the forbidden region. Approximating $\rho(E)$ deep in the forbidden region by the Airy density \\ $\rho(E) \propto \exp(-\frac{4 \sqrt{2}}{3}e^{S_0} |E|^{3/2})$, we obtain the saddle point $\bar E \propto - T^{-2} e^{-2 S_0}$. However, once $\bar E \sim - e^{-2 S_0/5}$ (which is equivalent to \eqref{s*}) the JT corrections to the WKB exponent in $\rho(E)$ become important. This leads to $T_* = \O(e^{-4S_0/5})$. This breakdown can alternatively be inferred from the rearrangement of the JT genus expansion as an expansion in powers of $T$ whose leading term is the Airy limit \cite{Okuyama}.

Finally, it is worth mentioning that there is an analog of the Airy limit for the Altland-Zirnbauer ensembles, where the Airy kernel is replaced by the Bessel kernel \cite{Nagao}. There exist analytic results for the distributions of the smallest eigenvalue \cite{TW_bessel} and the first gap \cite{Forrester} (at least when $\upbeta = 2$) in this case as well. It would be interesting to use these results to compute $F_q(0)$ and $\expect{S(T)}$ for JT supergravities that are nonperturbatively well-defined.

\section*{Acknowledgments}
We thank Raghu Mahajan and Douglas Stanford for useful discussions.

\appendix
\section*{Appendix: asymptotic behavior of the gap distribution}

\subsection*{A1 ~~ Perret-Schehr distribution} \label{app:d2}
In order to obtain the small $\tilde r$ behavior of $p_{\rm gap}(\tilde r|s)$ whose exact form is given in \eqref{cond2}, we use the expansion \cite{PS}
\be
f(\tilde r,s) = f(0,s) - \tilde r f_1(s) + \tilde r^2 f_2(s) +\O(\tilde r^3) \,,
\ee
where the first two coefficients are
\be
f(0,s) = q(s),\qquad f_1(s) = q'(s) + q(s) R(s) \,,
\ee
and $q,R$ were defined in Eqns. \eqref{qofs}-\eqref{pdfTW}. It follows from the form of \eqref{cond2} that $f_2(s)$ is not needed for the $\tilde r^2$ term in $p_{\rm gap}$. After some partial integrations, the coefficient $d_2(s)$ defined in \eqref{d2} is given by
\be
d_2(s) =
\frac{1}{R(s)}\left(\frac{1}{3} R^4(s) - q^2(s) R^2(s) + R(s) \int_s^\infty (q'^2(u) +q^4(u))\di u 
-\frac{1}{4}(R^2(s) -q^2(s))^2\right) \,.
\ee
Our interest is in the asymptotic behavior of $d_2(s)$. In the limit $s\gg 1$
\be\begin{split}\label{q,R}
	q(s) &= \frac{e^{-2 s^{3/2}/3}}{2\sqrt{\pi}s^{1/4}} \left( 1+\O(s^{-3/2}) \right) \,, \\
	R(s) &= \frac{e^{-4 s^{3/2}/3}}{8 \pi s} \left( 1+\O(s^{-3/2}) \right) \,,
\end{split}\ee
from which we get 
\begin{align}
	d_2(s) &\approx \frac{1}{R(s)}\left[\int_s^\infty q^2(u) \di u \int_s^\infty q'^2(u) \di u - 
\left(\int_s^\infty q(u) q'(u) \di u\right)^2\right] \notag \\
&= \frac{e^{-4s^{3/2}/3}}{128 \pi s^3} \left( 1 + \O(s^{-3/2}) \right) \,,\qquad s\to \infty \,. \label{d2right}
\end{align}
Note that even though the power corrections in \eqref{q,R} are naively relevant, they cancel in the leading term of $d_2(s)$.

In the opposite extreme, we can use the asymptotic behavior \cite{TW}
\be
R(s) = \frac{1}{4} s^2 +\O(s^{-1}) \,,
\ee
and $R'(s) = - q^2(s)$ to conclude that
\be
d_2(s) = \frac{s^6}{768}+\O(s^3) \,, \qquad s \to -\infty \,.
\ee

\subsection*{A2 ~~ WKB approximation} \label{app:wkb}
Exponentiating the Vandermonde determinant, the effective action for the eigenvalues of a unitary ensemble with potential $V(H)$ is
\be
I = N \sum_i V(\lambda_i)- \sum_{i< j} \log(\lambda_i - \lambda_j)^2 \,.
\ee
Suppose the classical upper edge of this distribution is at $a$. If the largest eigenvalue is fixed at $\lambda_{\rm max} =a+s$, the smaller eigenvalues can be thought of as eigenvalues of a new ensemble with potential 
\be\label{V1}
\tilde V(\lambda) = V(\lambda) - \frac{1}{N} \log(a+s-\lambda)^2 \,, \qquad -\infty < \lambda <a+s \,,
\ee
up to corrections that become irrelevant in the $N\to \infty$ limit. Therefore, finding $p_{\rm gap}(\tilde r|s)$ is equivalent to finding the distribution of the largest eigenvalue in this potential. While this is hard in general, far in the forbidden region, i.e. $s \gg 1$, we can approximate the distribution of the largest eigenvalue by the collective density of eigenvalues at fixed $s$, $\rho(x|s)$. This is in turn calculable via a WKB approximation (see \cite{Saad} for a derivation):
\be
\rho_{\rm WKB}(x|s) =\frac{1}{8 \pi x} \exp\left( 2 N \int_0^x \di x' y(x') \right) \,,
\ee
where $y$ is the spectral curve, given in terms of the tree level resolvent $R_{0,1}(X) \equiv \frac{1}{N} \expect{\frac{1}{X-H}}^{(0)}$ by
\be\label{y}
y(x) = R_{0,1}(X(x)) - \frac{1}{2} \tilde V'(X(x)),\qquad X(x) = a_+ +x \,,
\ee
and $a_+$ is the upper edge of the classical spectrum in $\tilde V$.

We follow the standard procedure for finding $R_{0,1}$ and the classical edges. First, the saddle-point equation gives
\be
R_{0,1}(X+i\ep) + R_{0,1}(X-i\ep) = \tilde V'(X|s) \,.
\ee
In a convex potential, one can use Cauchy's theorem to derive the following expression for $R_{0,1}$ as an integral along a single cut between the two classical edges of the spectrum $a_{\pm}$:
\be\label{r01}
R_{0,1}(X) = \int_{a_-}^{a_+} \frac{\di \lambda}{2\pi i} \frac{\tilde V'(\lambda)}{\lambda - X}\sqrt{\frac{\sigma(X)}{\sigma(\lambda)}}
\ee
where 
\be
\sigma(X) =(X- a_+)(X-a_-) \,.
\ee
The endpoints are fixed by
\be\label{c1}
\int_{a_-}^{a_+} \frac{\di \lambda \, \tilde V'(\lambda)}{\sqrt{\sigma(\lambda)}} = 0 \,, \qquad
\int_{a_-}^{a_+} \frac{\di \lambda}{i \pi}\frac{\lambda \, \tilde V'(\lambda)}{\sqrt{\sigma(\lambda)}} =-2 \,.
\ee
Note that in order to regard $R(X)$ as a complex function and apply the above manipulations, it is necessary to map the upper limit of $\lambda$ in \eqref{V1} to $\infty$, so that we have a potential that is defined on the entire real axis. This can be done via the change of variable 
\be
X_* = X - X_0 \log\frac{a+s - X}{X_1} \,,
\ee
for any positive $X_0$ and $X_1$. Since $X_0$ is arbitrary, by taking it to be much smaller than any relevant scale, we can work with the original variable $X$ as long as we are computing quantities that do not diverge as $X\to a+s$. 

The Airy limit, with spectral curve $\sqrt{x}$, can be obtained by taking the original potential to be a Gaussian, with a particular scaling of the width with $N$. With this choice
\be\label{v'}
\tilde V'(X) = \frac{4}{a^2} X +\frac{2}{N (\lambda_{\rm max}-X)},\qquad a = 2 N^{2/3} \,.
\ee
We are keeping $s$ finite as $N$ and $a$ are sent to infinity. Keeping terms that remain finite in this limit, we obtain from \eqref{c1}
\be
a_- = -a \,,
\ee
and the relation
\be\label{delta}
\delta - \frac{2}{\sqrt{\delta}} = s,\qquad \delta \equiv a+s - a_+ \,.
\ee
In the limit $s\gg 1$ (corresponding to $\lambda_{\rm max}$ moving far in the forbidden region of the original potential), $a_+$ relaxes to $a$:
\be\label{delta2}
\delta = s + \frac{2}{\sqrt{s}} + \O(s^{-2}) \,, \qquad s \gg 1 \,.
\ee
On the other hand, when $s\ll -1$ we have $\delta \ll 1$, implying that the classical edge is pushed close to $\lambda_{\rm max}$. 

Next we calculate $R_{0,1}$ for the specific potential \eqref{v'} using \eqref{r01}. There are two contributions, one from the original potential and the other from the repulsion of $\lambda_{\rm max}$. They respectively are proportional to 
\be
\int_{a_-}^{a_+} \di \lambda \frac{\lambda}{(\lambda-X) \sqrt{(\lambda-a_-)(a_+-\lambda)}} 
= \pi\left(1-\frac{X}{\sqrt{\sigma(X)}}\right) \,,
\ee
and 
\be
\int_{a_-}^{a_+} \frac{\di \lambda }{(\lambda-X)(\lambda-\lambda_{\rm max}) \sqrt{(\lambda-a_-)(a_+-\lambda)}} 
=\frac{\pi}{\lambda_{\rm max} - X} \left(\frac{1}{\sqrt{\sigma(X)}}-\frac{1}{\sqrt{\sigma(\lambda_{\rm max})}}\right) \,.
\ee
They result in
\be\label{R01ex}
R_{0,1}(X) = \frac{2}{a^2} (X - \sqrt{\sigma(X)}) +
\frac{1}{N(\lambda_{\rm max} - X)} \left(1 -\sqrt{\frac{\sigma(X)}{\sigma(\lambda_{\rm max})}}\right) \,.
\ee
Substituting this solution in \eqref{y}, and taking the double-scaling limit, gives
\be
N y(x) = -\sqrt{x}- \frac{1}{\delta -x}+ \frac{1}{\delta + \sqrt{\delta x}}  \,,
\ee
from which we obtain
\be\label{p10}
\rho_{\rm WKB}(x|s) =
\frac{1}{8\pi x}\left(1-\frac{x}{\delta}\right)^2
\exp\left(-\frac{4}{3} x^{3/2}+4\left(\sqrt{x/\delta}-\log(1+\sqrt{x/\delta})\right)\right) \,,
\ee
where $\delta(s)$, the distance between $\lambda_{\rm max}$ and $a_+$ is determined from \eqref{delta}. As argued above, when $x\to \delta$ we are approaching $\lambda_{\rm max}$ and the density of the $N-1$ eigenvalues with $\lambda_{\rm max}$ fixed, $\rho(x|s)$, is a good approximation to $p_{\rm gap}(\tilde r|s)$, where $\tilde r = \delta -x$. The WKB approximation to $\rho(x|s)$ is good when $s\to \infty$, as can be seen by noticing that loop corrections to the resolvent lead to an expansion in inverse powers of $x$ and $\delta$ in the WKB exponent. It follows from \eqref{delta} that neglecting those corrections is justified when $s\gg 1$. Substituting \eqref{delta2} in \eqref{p10}, expanding to $\O(\tilde r^3)$ and focusing on the leading large-$s$ behavior, we get
\be
\rho_{\rm WKB}(\tilde r|s) ~ \underset{\tilde{r}\to 0}= ~ \frac{e^{-\frac{4}{3} s^{3/2}}}{128\pi s^3} \tilde r^2 
\left( 1+ 2 s^{1/2} \tilde r + \mathcal{O}(\tilde{r}^2) \right) \,,
\ee
which agrees with \eqref{d2right}. It also shows that $p_{\rm gap}(\tilde r|s)$ is not even in $\tilde r$.

\bibliographystyle{utphys}
\bibliography{refs}
\end{document}